\documentclass[twocolumn,prx,aps,showpacs,superscriptaddress,floatfix,nofootinbib,10pt]{revtex4}

\usepackage{amsfonts,amsmath,amssymb}
\usepackage{graphicx}
\usepackage{color}
\usepackage[normalem]{ulem}
\usepackage{amsfonts}
\usepackage[toc,page]{appendix}
\usepackage[ocgcolorlinks,colorlinks=true,linkcolor=blue,citecolor=red]{hyperref}
\usepackage{latexsym}
\usepackage{amsfonts}
\usepackage{algpseudocode}
\usepackage{amsthm}
\usepackage{mathrsfs}
\usepackage{natbib}
\usepackage{color,verbatim}
\usepackage{psfrag}

\DeclareMathOperator{\Tr}{tr}

\newcommand{\be}{\begin{eqnarray}}
\newcommand{\ee}{\end{eqnarray}}

\begin{document}
\title{Quantum communication games manifest preparation contextuality}

\author{Armin Tavakoli}
\affiliation{Department of Physics, Stockholm University, S-10691 Stockholm, Sweden}
\affiliation{Groupe de Physique Appliqu\'ee, Universit\'e de Gen\'eve, CH-1211 Gen\'eve, Switzerland}


\date{\today}


\begin{abstract}
Communication games are collaborative information processing tasks involving a number of players with limited communication. Such games are useful tools for studying physical theories. A physical theory exhibits preparation contextuality whenever observed behaviour cannot be explained by a preparation noncontextual model. Here we show that there is a fundamental connection between the performance in communication games and the degree of preparation (non)contextuality. For this purpose, we present a general framework that allows us to construct communication games such that the game performance corresponds to a measure of preparation (non)contextuality.
We illustrate the power of this framework by, 1) deriving many examples of tests of preparation contextuality, 2) showing that quantum violations of Bell inequalities can be derived from  quantum violations of preparation noncontextuality, 3) qualitatively and quantitatively explaining related previous results on preparation contextuality in quantum communication games, and 4) solving the open problem of revealing the preparation contextuality of the maximally mixed quantum state in any dimension. 
	
\end{abstract}



\maketitle

\section{Introduction}
Communication games are tools by which one can study fundamental limiting features of physical theories in terms of their ability to process information. In these games, a number of parties intend to jointly solve a task despite the amount and type of communication being constrained by some rules. Therefore, the partnership can only solve the task with some probability. The probability of solving the task depends on the physical theory by which the partnership is assumed to operate. For this reason, communication games are often used to study the limitations of physical theories \cite{IC09, GH14, BB15}, in particular the relation between quantum theory and classical theories. Therefore, these games provide tools for identifying and quantifying quantum supremacy \cite{GB10, HG12, AB12, PB11, LP12, AB14}.

Interestingly, examples of communication games are known in which the better-than-classical performance in the game constitutes a certificate of the system exhibiting preparation contextuality \cite{SB09, A16}. Preparation noncontextuality is the assumption that, for a given operational theory, if two preparations of a system cannot be distinguished by any measurement allowed in that theory, then the respective hidden variables which determine the ontic states of the two systems must have the same distribution \cite{Sp05}. If this assumption can be falsified for an operational theory, then that theory manifests preparation contextuality. Preparation contextuality (and also the analog concept of measurement contextuality \cite{Sp05}), has been applied in several interesting ways \cite{SB09, LM13, CK16, KS15, A16} to study foundational physical problems. 

The above motivates a much broader question: is there a general connection between outperforming classical limitations in communication games and preparation contextuality? Here, we will present a general framework for constructing communication games in such a way that the performance in a game corresponds to the value of the operator in a preparation noncontextuality inequality, i.e., an inequality satisfied by all preparation noncontextual models, the violation of which implies that the system manifests preparation contextuality. We will provide explicit examples in which we derive and violate such preparation noncontextuality inequalities with quantum theory. Importantly, we will show that the amount to which quantum theory is able to violate a Bell inequality can be derived from the ability of quantum theory to violate a preparation noncontextuality inequality. This allows us to understand the Tsirelson bound of Bell inequalities as a limitation following from the degree of preparation contextuality allowed in quantum theory. Furthermore, using this strong connection between preparation contextuality and Bell inequality violations in quantum theory, we can qualitatively and quantitatively explain the degree of preparation contextuality observed in Refs.\cite{SB09, A16}, and also solve an open problem in this field; namely demonstrating the preparation contextuality of the maximally mixed quantum state in any dimension.

\section{Communication games} \label{1}

In a two-player communication game, a party Alice (Bob) holds a set of data denoted $x\in I_A$ ($y\in I_B$) sampled with a probability distribution $p_{A}(x)$ ($p_B(y)$)  over the space $I_A$ ($I_B$). 

Alice will encode $x$ into a information-carrying resource (a preparation) which is sent to Bob who attempts to decode it according to a measurement labeled $y$. This yields an outcome $b$ from which Bob aims to compute some task-functions $\{T_k(x,y)\}_{k=1}^{N}$ for his chosen $y$. We require that the task-functions are such that $\forall x\forall y: T_k(x,y)\neq T_{k'}(x,y)$ for $k\neq k'$ i.e. Bob can at most compute one task-function for a given outcome $b$. Depending on which task-function (if any) Bob manages to compute with $b$, the partnership receives a payoff $\$_k(x,y)$. For simplicity and without loss of generality one can normalize the payoffs such that $\left|\$_k(x,y)\right|\leq 1$. The average payoff earned by the partnership is written
\begin{equation}\label{payoff}
	\langle \$\rangle_{x,y}\!=\!\!\sum_{x\in I_A}\sum_{y\in I_B} p_{A}(x)p_B(y)\sum_{k=1}^{N}\$_k(x,y)p(b=T_k|x,y).
\end{equation} 
We call this the \textit{performance} of the game.

So far, Alice could just send Bob her entire data set $x$ from which he could trivially compute the most rewarding $T_k$ and achieve maximal performance in the game. Therefore, we must impose some set of communication constraints which, at the very least, should forbid trivial strategies and ensure that the communication game is nontrivial.

\section{Communication games as tests of preparation contextuality} \label{s2}
A model is said to be preparation noncontextual \cite{Sp05} if
\begin{equation}\label{noncontex}
	\forall y\forall b: p(b|x,y)=p(b|x',y)\Rightarrow p(\lambda|x)=p(\lambda|x'),
\end{equation}
where $\lambda$ is the hidden variable, $x$ and $x'$ are associated to two preparations and $y$ is associated to a measurement. That is, if there is no measurement $y$ that can distinguish between $x$ and $x'$, then they are assumed to be associated to the same hidden variable distribution.  

We will now show that one can systematically choose suitable communication constraints for communication games, in such a way that the premise of Eq.\eqref{noncontex} is satisfied and that the assumption of preparation noncontextuality leads to a preparation noncontextuality inequality in which the performance of the game is the operator.

Partition $I_A$ in $L$ different ways, each into $D$ sets labeled $S_i^j$ with $i\in\{1,\ldots,D\}$ and $j\in\{1,\ldots, L\}$. That is; $\forall j: \bigcup_{i=1}^{D}S_i^j=I_A $ while $\forall j: S_i^j\cap S_{i'}^j=\emptyset$  if $i\neq i'$.

Now, choose communication constraints as follows: impose an \textit{obliviousness} constraint
\begin{equation}\label{obliv}
\forall y\forall b: \frac{1}{q_{i,j}}\sum_{x\in S_i^j} p(x|b,y)=\frac{1}{q_{i',j'}}\sum_{x\in S_{i'}^{j'}} p(x|b,y).
\end{equation} 
Here $q_{i,j}=p(x\in S_i^j)=\sum_{x\in S_i^j}p_A(x)$ serves as a normalization. Thus, no matter the performed measurement and observed outcome; Bob gains no information, as compared to what he knew before communication, about to which set $S_i^j$  the data $x$ of Alice belongs.  To see that this in fact leads to the premise of the preparation noncontextuality assumption in Eq. \eqref{noncontex}, we apply Bayes' rule to the above summands: $p(x|b,y)=p(b|x,y)p(x|y)/p(b|y)$. Using that preparations are independent of how they are to be measured, we can write the obliviousness constraint in Eq.\eqref{obliv} as
 \begin{equation}\label{obliv2}
 \forall y\forall b:\sum_{x\in S_i^j}p(b|x,y)\frac{ p_A(x)}{q_{i,j}}=\sum_{x\in S_{i'}^{j'}}p(b|x,y)\frac{p_A(x)}{q_{i',j'}}.
 \end{equation}  
 We note that $\{p_A(x)/q_{i,j}\}_{x\in S_i^j}$ is a valid probability distribution normalized over the set $S_i^j$. Therefore, each side of Eq.\eqref{obliv2} can be regarded as a convex combination.

 Now, note that the probability that the outcome $b$ for a given measurement was obtained from a measurement on a preparation associated to $S_i^j$ is the convex mixing of its constitutes: $p(b|x\in S_i^j,y)=
 \sum_{x\in S_i^j} p(b|x,y)p_A(x)/q_{i,j}$. By the same token, the distribution of the hidden variable is $p(\lambda|x\in S_i^j)=\sum_{x\in S_i^j} p(\lambda|x)p_A(x)/q_{i,j}$. This, together with Eq.\eqref{obliv2}, implies that $\forall y\forall b: p(b|x\in S_i^j,y)=p(b|x\in S_{i'}^{j'},y)$. This is exactly the form of the premise of the preparation noncontextuality statement in Eq. \eqref{noncontex}. Thus, a preparation noncontextual hidden variable model would require that $p(\lambda|x\in S_i^j)=p(\lambda|x\in S_{i'}^{j'})$. Using Bayes' rule we find that $p(x\in S_i^j|\lambda)/q_{i,j}=p(x\in S_{i'}^{j'}|\lambda)/q_{i',j'}$ which means that even with the knowledge of the hidden variable the obliviousness constraint still holds. 

The obliviousness constraint limits the composition of Alice's preparations. For such tasks, there exists a computable optimal preparation noncontextual hidden variable performance $p^{pnchv}$ obtained by maximizing $\langle \$\rangle_{x,y}$ over all deterministic strategies that respect the obliviousness constraint in Eq.\eqref{obliv}. Hence,
\begin{equation}
\langle \$\rangle_{x,y} \leq p^{pnchv}
\end{equation}
is a preparation noncontextuality inequality.

Clearly, there is a plethora of ways in which one can choose the communication constraints for a given communication game so that the performance corresponds to a preparation noncontextuality inequality. However, it is not obvious which of our preparation noncontextuality inequalities that can be violated in quantum theory. Nevertheless, we can find many cases in which such violations are possible. We begin the exploration of preparation contextuality in quantum theory by providing a family of examples.

\subsection{Example: Parity-oblivious random access codes}\label{EX1}
We consider a broad family of communication games, known as random access codes \cite{T15},  in which Alice holds $x=x_1\ldots x_{n}\in\{0,\ldots,d-1\}^n=I_A$ with $p_A(x)=1/d^n$, and Bob holds $y\in\{1,\ldots,n\}$ with $p_B(y)=1/n$. We require only a single task-function, namely $T=x_y$ with an associated payoff $\$=1$. We partition $I_A$: for every $j=j_1\ldots j_n\in\{0,1\}^n$ with $\sum_{r=1}^nj_r\geq 2$, partition $I_A$ into $S_i^j=\{x|\sum_{r=1}^{n}x_rj_r=i\mod{d}\}$ for $i=0,\ldots, d-1$ and impose the corresponding obliviousness constraint Eq.\eqref{obliv}. The obliviousness constraint is interpreted as Bob not being allowed to gain any information on the modulo $d$ sum (parity) of any string of Alice's data with at least two elements. Note that in this case $\forall i\forall j: q_{i,j}=1/d$ so these cancel in Eq.\eqref{obliv}. The preparation noncontextuality inequality for this family of parity-oblivious random access codes reads 
\begin{equation}\label{ex1}
\langle \$\rangle_{x,y}=\frac{1}{nd^n}\sum_{y=1}^{n}\sum_{x\in \{0,\ldots,d-1\}^n} p(b=x_y|x,y)\leq \frac{n+d-1}{nd}.
\end{equation}
The right-hand-side is the preparation noncontextual bound. To compute this, note that the obliviousness constraint is invariant under permutations of the order of the elements $x_1\ldots x_n$ in Alice's string $x$, and if we send more than one entry of the string $x$  the obliviousness constraint is violated. Therefore, we can without loss of generality imagine that Alice always sends her first entry $x_1$ to Bob. Hence, if $y=1$ Bob always finds $b=x_y$, whereas if $y\neq 1$ he will have to guess, succeeding with probability $1/d$. Calculating this average returns the bound in Eq.\eqref{ex1}. 

The preparation noncontextuality inequalities derived in Ref.\cite{SB09} and Ref.\cite{A16} constitute special cases of the above corresponding to us setting $d=2$ and $n=2$ respectively. Naturally, we may consider choices of $(n,d)$ beyond these special cases. For sake of examplification, we have numerically optimized the the case of $(n,d)=(3,3)$ in a quantum model using a see-saw method\footnote{In a quantum model, the objective function in Eq.\eqref{ex1} is linear in the communicated quantum states and the measurements of Bob respectively. For fixed states, the optimization problem is a semidefinite program over the measurements, and vice versa. In a see-saw approach, we first optimize over states for fixed measurements, and then optimize over measurements for fixed states etc. until the value of objective function converges to a satisfactory precision. This provides a lower bound on the left-hand-side of Eq.\eqref{ex1} in a quantum model.} with semidefinite programs and obtained $\langle \$\rangle_{x,y}^Q\approx 0.6711$ violating the preparation noncontextual bound $\langle \$\rangle_{x,y}^{pnchv}=5/9$.

\subsection{Bell-type quantum correlations are preparation contextual}\label{II}
Here, we shall show that under the assumption of quantum theory, the ability and extent to which a Bell inequality can be violated can be understood from the ability of quantum theory to manifest preparation contextuality.  

We consider general bipartite Bell inequalities, with $m_A$ ($m_B$) settings for Alice (Bob) and $d$ outcomes, on the form introduced in Ref.\cite{TZ16}. These Bell inequalities read
\begin{multline}\label{B}
\sum_{x,y}p_A(x)p_B(y)\sum_{i=1}^{N}\sum_{k=0}^{K}\$_{xy}(i,k)P_{xy}(a+b=F^i_{xy}(k))\\\leq B,
\end{multline}
where $B$ is the classical bound, $F_{xy}^i$ are some functions onto the set $\{0,\ldots, d-1\}$ such that $\forall x,y:$ the ranges of $F_{xy}^j$ and $F_{xy}^l$ are disjoint for $j\neq l$. Also, $N$ and $K$ are some natural numbers, and $a+b$ is computed modulo $d$. A multitude of known Bell inequalities can be written on this general form.

Communication games were introduced in Ref.\cite{TZ16} based on the above Bell inequalities. In these games Alice is given $x_0\in\{0,\ldots d-1\}$ with $p(x_0)=1/d$. In addition Alice and Bob hold $x\in\{0,\ldots,m_A-1\}$ and $y\in\{0,\ldots,m_B-1\}$ respectively with associated distributions $p_A(x)$ and $p_B(y)$. Bob aims to compute the functions $T_{i,k}(x_0,x,y)=x_0+F_{xy}^i(k)\mod{d}$ which, if successful, returns the payoff $\$_{xy}(i,k)$. Alice will send Bob a message $M$ which he will use to calculate his guess $G_y\in\{0,\ldots,d-1\}$ for the value of $T_{i,k}$. The average payoff in this game is 
\begin{multline}\label{ccp2}
I=\frac{1}{d}\sum_{x_0=0}^{d-1}\sum_{x,y}p_A(x)p_B(y)\\
\sum_{i=1}^{N}\sum_{k=0}^{K}\$_{xy}(i,k)P(G_y=T_{i,k}(x_0,x,y)).
\end{multline}

So far, we have not introduced any communication constraints. In order to make the connection to preparation noncontextuality inequalities, we choose these as obliviousness constraints. To this end, we define $S_i=\{x_0x|x=i\}$ and require that Bob gains no information about to which $S_i$ the data $(x_0,x)$ of Alice belongs. In quantum theory, we write this as  
\begin{equation}\label{bell}
\sum_{x_0=0}^{d-1}\rho_{x_00}=\ldots=\sum_{x_0=0}^{d-1}\rho_{x_0(m_A-1)}.
\end{equation} 

In Appendix \ref{A} we show that in any preparation noncontextual encoding strategy of Alice, if she communicates more than $\log d$ bits about her data $(x_0,x)$, the obliviousness constraint will be violated. Using this knowledge, we let Alice communicate up to $\log d$ bits while respecting the obliviousness constraint. Such encodings are written  $M=x_0+a(x)\mod{d}$ for some function $a$. Here, $x_0$ completely randomizes the message so that no information about $x$ can be extracted from it\footnote{In fact, the message $M=x_0+a(x)\mod{d}$ is known to be optimal even when we only require at most $\log d$ bits of communication, without additionally demanding obliviousness of $x$ \cite{TZ16}.}. Alice sends the message to Bob who outputs the guess $G_y=x_0+a(x)+b(y)\mod{d}$. Inserting this guess into Eq.\eqref{ccp2}, one finds that $I$ becomes the equivalent of the left-hand-side of Eq.\eqref{B} and must therefore admit the bound 
\begin{equation}\label{prp}
I\leq B.
\end{equation}

Now, we make a limiting assumption: the marginal probability distributions of Alice in the Bell inequality are uniform. Given this assumpion, we will now show that in quantum theory, the values of $I$ are precisely those achievable by quantum correlations in the Bell inequality in Eq.\eqref{B}.

In a quantum approach to the communication game, Alice associates her inputs $(x_0,x)$ to the states $\rho_{x_0x}\in \mathbb{C}^D$ for some dimension $D$ such that Eq.\eqref{bell} is satisfied. Let Alice prepare some entangled state $\rho^{AB}\in\mathbb{C}^{D'}\otimes \mathbb{C}^D$ for any $D'$ of her choice. If the dimension $D'$ of one Hilbert space is sufficiently large, Neumark's theorem implies that there exists $m_A$ projective measurements indexed by $x$, with $d$ outcomes indexed by $x_0$, associated to measurement operators $A_{-x_0}^{x}$\footnote{The minus sign is introduced only for future convenience and serves as a local relabeling.}, such that $\rho_{x_0x}=d\Tr_A\left(A_{-x_0}^{x}\otimes \mathbf{1} \rho^{AB}\right)\in \mathbb{C}^D$  (note that we have made explicit use of the assumption of uniform marginals). That is; a set of preparations  $\{\rho_{x_0x}\}\in\mathbb{C}^D$ can be prepared by performing  measurements on a state of sufficiently large local Hilbert space dimension. Alice may then send her second subsystem to Bob, knowing that the communication constraints in Eq.\eqref{prp} are satisfied. The fact that the obliviousness constraint is satisfied follows from the assumption of uniform marginals:
\begin{equation}\label{const}
\forall x:	\frac{1}{d}\sum_{x_0=0}^{d-1}\rho_{x_0x}=\Tr_A\left(\sum_{x_0=0}^{d-1}A_{-x_0}^{x}\otimes \mathbf{1} \rho^{AB}\right)=\rho^B,
\end{equation}
which is independent of $x$.

If the measurement operators of Bob are labeled $B_{b}^y$ for $b=0,\ldots,d-1$, we can write the performance in any such communication game as 
\begin{multline}\label{res}
I=\sum_{x,y}p_A(x)p_B(y)\\
\sum_{i=1}^{N}\sum_{k=0}^{K}\$_{xy}(i,k)\sum_{x_0=0}^{d-1}\Tr\left(A_{-x_0}^{x}\otimes B_{T_{i,k}}^y \rho^{AB}\right)\\
=\sum_{x,y}p_A(x)p_B(y)\sum_{i=1}^{N}\sum_{k=0}^{K}\$_{xy}(i,k)P^Q_{xy}(a+b=F^i_{xy}(k)),
\end{multline}
where we in the second step have relabled Alice's and Bob's outcomes as $a$ and $b$ respectively and used that $-x_0+T_{i,k}=-x_0+x_0+F_{xy}^i(k)=F_{xy}^i(k)$. Also, $P^Q$ denotes conditional probability in a quantum model. The equation \eqref{res} is precisely the expression on of the left-hand-side of Eq.\eqref{B} in a quantum model. Thus, quantum correlations in tests of  local realism can be viewed as manifestations of preparation contextuality. Importantly, for a Bell inequality of the considered class that achieves its Tsirelson bound with a quantum probability distribution that has uniform marginals, we may understand the Tsirelson bound as a limitation imposed by degree of preparation contextuality allowed in quantum theory.

We emphasize that only a sub-class of preparation noncontextuality inequalities are relevant for Bell inequalities. It is easy to imagine explicit preparation noncontextuality inequaltities that do not admit the form considered in this section. For instance, such examples were discussed in section \ref{EX1}. Also, it is appropriately noted that in Ref.\cite{BB15} the Tsirelson bound of the  CHSH inequality \cite{CHSH69} has been connected to a the degree of preparation contextuality allowed in quantum theory using a particular game outlined in Ref.\cite{SB09}. Our results in this section generalize this connection to arbitrary Bell inequalities.   

\subsection{Explaining the preparation contextuality under parity-obliviousness in Refs.\cite{SB09, A16} }

In section \ref{EX1} we mentioned that the preparation noncontextuality inequalities of Refs.\cite{SB09, A16} correspond to special cases of the family of preparation noncontextuality inequalities considered in that section. These were based on random access codes with an additional parity oblivious constraint. 

It is interesting to point out that the optimal quantum performance when $(n,d)=(2,2)$ is no different \cite{SB09} than what is found for the standard $2\rightarrow 1$ random access code \cite{W83, A99} in which the parity-oblivious constraint is relaxed to Alice sending at most one bit of information to Bob. The additional parity-oblivious constraint just happens to be satisfied by the optimal quantum protocol for the $2\rightarrow 1$ random access code. Nevertheless, parity-obliviousness is the reason that the link to preparation contextuality can be made. However, it was shown that this equivalence does not hold for higher-dimensional cases \cite{A16}: when $d=3,4,5$ the optimal (known) quantum performance is lower than the analog quantum random access code \cite{T15} i.e., the parity-oblivious constraint has a non-trivial impact on the quantum performance. 

We will now apply the results of the previous section to explain qualitatively the difference between the $d=2$ case as compared to $d>2$, and quantitavely explain the amount of preparation contextuality observed in Refs.\cite{SB09, A16}. First, notice that when we set $n=2$ the preparation noncontextual bound in Eq.\eqref{ex1} becomes $p^{pnchv}=1/2(1+1/d)$, which coincides with the classical bound of the associated random access code with two $d$-valued inputs in Ref.\cite{T15}. Random access codes with $n=2$ have a realization based on  entanglement powered by the violation of a Bell inequality in which Alice (Bob) has $d$ ($2$) settings with $d$ outcomes \cite{TM16}. These inequalities take the form of Eq.\eqref{B}. For $d=3,4,5$ the maximal known violations of these inequalities are obtained from measurements on a maximally entangled state of local dimension $d$. Therefore, the marginal distributions of Alice and Bob are necessarily uniform, independent of the choice of measurements. Thus, from the discussion in the previous section, we expect that the maximal quantum preparation contextuality coincides with the Tsirelson bound of these Bell inequalities. This falls well in line with the optimizations performed in Ref.\cite{A16} for $d=3,4,5$  returning that the optimal quantum performance is found when the communicated state of Alice, averaged over the partitions of Alice's input space, is the maximally mixed state in dimension $d$. Indeed, the known maximal violations of the Bell inequalities tailored to random access codes, presented in Ref.\cite{TM16}, do coincide very accurately with the results of Ref.\cite{A16} for $d=3,4$. For $d=5$ the violation of the Bell inequality is even slightly larger than what was numerically obtained in Ref.\cite{A16}, which means that quantum preparation contextuality is in fact slightly stronger than initially believed. 

In contrast, when we set $d=2$, the Bell inequality in Ref.\cite{TM16} reduces to the CHSH-inequality, for which the Tsirelson bound is $p^Q=1/2\left(1+1/\sqrt{2}\right)$ \cite{Tsi80}. This is the same as the quantum performance of the $2\rightarrow 1$ random access code with quantum communication. Therefore, we would not expect parity-obliviousness to have any influence on the maximal quantum preparation contextuality for the game of Ref.\cite{SB09} corresponding to $(n,d)=(2,2)$ in section \ref{EX1}.

\subsection{Revealing the preparation contextuality of the maximally mixed state}

Spekkens showed the preparation contextuality of the maximally mixed qubit state \cite{Sp05}, and a similar proof for arbitrary mixed qubit states was given in Ref.\cite{BB14}. By use of communication games, the preparation contextuality of the maximally mixed quantum state  of dimension $d=3,4,5$ was shown in Ref.\cite{A16}, and the extension to any $d$ was left as the main open problem. We will now see that due to our connection between preparation contextuality in quantum theory and quantum correlations violating a Bell inequality, this open problem can be straightforwardly solved. 

To this end, we consider the CGLMP inequality \cite{CGLMP02}, which is a Bell inequality with two settings and $d$ outcomes constituting a face of the local polytope. This particular inequality has been cast as a communication game in Ref.\cite{magic7} with quantitative communication constraints. Also, the CGLMP inequality admits the form of Eq.\eqref{B} and is therefore a special case of our discussion in section \ref{II}.   

The game we associate to the CGLMP inequality is as follows. Alice holds $x=x_0x\in\{0,\ldots,d-1\}\times \{0,1\}=I_A$ with $p_A(x)=1/2d$, and Bob holds $y\in\{0,1\}$ with $p_B(y)=1/2$. There are $2\lfloor \frac{d}{2}\rfloor$ different task-functions, which we write as $T_k^q=x_0-(-1)^{x+y+q}(k+q)-xy\mod{d}$ for $k=0,\ldots,\lfloor \frac{d}{2}\rfloor-1$ and $q=0,1$. The payoff function for computing $T_k^q$ is $\$_{k,q}=(-1)^q\left(1-\frac{2k}{d-1}\right)$. Alice's communication must satisfy an obliviousness constraint which in quantum theory reads $\sum_{x_0=0}^{d-1}\rho_{x_00}=\sum_{x_0=0}^{d-1}\rho_{x_01}$.

By our results in section \ref{II}, the classical bound of the CGLMP inequality also bounds the performance of the described game:
\begin{multline} \label{ex2}
\langle \$\rangle_{x,y}=\frac{1}{4d}\sum_{x,y=0,1}\sum_{x_0=0}^{d-1} \Bigg[\sum_{k=0}^{\lfloor \frac{d}{2}\rfloor-1}\left(1-\frac{2k}{d-1}\right)\\
\left(p(b=T_k^0|x_0,x,y)-p(b=T_k^1|x_0,x,y)\right)\Bigg] \leq 1/2.
\end{multline}
In fact, the upper bound can be achieved by a preparation noncontextual strategy simply by Alice always sending $x_0$ to Bob, and Bob always outputting $b=x_0$. 

The CGLMP inequality can be violated for any number of outcomes $d$ by performing measurements on a shared state of local Hilbert space dimension $d$ \cite{ZG08}. The shared entangled state leading to an optimal quantum violation is not maximally entangled. However, one can achieve quantum violations of the CGLMP inequality for any $d$, although not optimal, by Alice and Bob sharing the maximally entangled state with measurements that give rise to a quantum probability distribution with uniform marginals \cite{CGLMP02, ZG08}. Using the arguments of the previous section, we can immediately associate states and measurements in our above communication game that violate the constraints of preparation noncontextuality for the maximally mixed state in dimension $d$. We explicitly calculate the violation of Eq.\eqref{ex2} by preparations averaging to the maximally mixed quantum state in Appendix \ref{appB}.

\section{Conclusions}
Our two main results were: i) a general framework for imposing a communication constraint on arbitrary bipartite communication games so that they can be mapped to measures of preparation (non)contextuality, and ii) that quantum correlatins violating a Bell inequality can be understood as special instances of preparation contextuality. In particular, we used the later to first provide a simple explanation of the results of Refs.\cite{SB09, A16}, and then to solve the open problem of revealing the preparation contextuality of the maximally mixed quantum state in any dimension.

There are several key open problems and directions of future research of which we mention a few: 1) Do communication games with communication constraints that do not respect any obliviousness constraint (and hence do not map to preparation noncontextuality inequalities) admit a connection to some fundamental physical assumption in the same spirit as presented here for games respecting an obliviousness constraint? 2) If a communication game involves more than two players, can a general connection similar to the one outlined here be established between the performance of the game and the operator in a preparation and transformation noncontextuality inequality? 3) Our games require the obliviousness constraint to be satisfied by the sender, whilst making no assumptions on the measurements of the receiver. Therefore, it would be interesting if one can find one-sided device independent quantum information applications powered by preparation contextuality.

\section{Acknowledgements}
The authors thank Ad\'an Cabello for criticism and insightful comments. The author acknowledges financial support from the Swiss National Science Foundation (Starting grant DIAQ).

\appendix

\section{Upper bound on classical Communication for games in section \ref{II}}\label{A}

Here, we prove a limitation of preparation noncontextual strategies for the communication games considered in section \ref{II}. Explicitly, we show that in a classical picture in which information is encoded into integer values, communication of more than $\log d$ bits of information about Alice's input $(x_0,x)$ violates the obliviousness constraint in Eq.\eqref{bell}. The proof is a straightforward modification of the proof presented in Ref.\cite{A16}.

Alice encodes her inputs $x_0,x\in\{0,\ldots,d-1\}\times \{0,\ldots,m_A-1\}=I_A$ using a classical encoding function $E:\{0,\ldots,d-1\}\times \{0,\ldots,m-1\}\rightarrow \{0,\ldots,L\}$ for some $L$. This corresponds to a partition of Alice's input space $I_A$ into $L+1$ non-empty sets labeled $R_j$ for $j=0,\ldots,L$. We remind ourselves that Alice's data is partitioned into $S_i=\{x_0x|x=i\}$. If Bob is to gain no information, as compared to what was already known to him due to the distribution $p_A(x)$ of Alice's input, the following must be satisfied:
\begin{equation}\label{req}
	\forall r,t,j: \left|R_{j}\cap S_r\right|=\left|R_{j}\cap S_{t}\right|.
\end{equation} 
If we assume that $L\geq d$, then by the pigeonhole priciple, there exists at least one $j^*\in\{0,\ldots,L\}$ such that $|R_{j^*}|< m$. Thus, there exist at least one $l^*$ such that $\left|R_{j^*}\cap S_{l^*}\right|=0$ which contradicts the requirement in Eq.\eqref{req}. Thus, the obliviousness constraint cannot be satisfied when $L\geq d$.

\section{Certifying the preparation contextuality of maximally mixed quantum state}\label{appB}

In this appendix we calculate the violation of Eq.\eqref{ex2} obtained for the maximally mixed quantum state.

Let Alice and Bob share some $d\times d$ entangled state written on Schmidt form as $|\psi\rangle=\frac{1}{\sqrt{N}}\sum_{k=0}^{d-1}\gamma_k |kk\rangle$ in which $\gamma_k\in \mathbb{R}$ and $N=\sum_{k=0}^{d-1}\gamma_k^2$. The associated density matrix is $\rho^{AB}$. Alice has two measurement options, indexed by $x\in\{0,1\}$, given by $|x_0\rangle_{{x},A}=\frac{1}{\sqrt{d}}\sum_{k=0}^{d-1}\omega^{k(x_0+\alpha_{x})}|k\rangle$ for $x_0=0,\ldots, d-1$ with $\alpha_0=0$ and $\alpha_1=1/2$. The associated projection operator is $A_{x_0}^{x}$. 

Let Bob perform the measurements used to maximally violate the CGLMP inequalities: $|b\rangle_{{y},B}=\frac{1}{\sqrt{d}}\sum_{k=0}^{d-1}\omega^{k(-b+\beta_{y})}|k\rangle$ for $b=0,\ldots, d-1$, $y\in\{0,1\}$ and $\beta_y=(-1)^y1/4$. The associated projection operator is written $B_b^y=|b\rangle_{{y},B}\langle b|_{{y},B}$. Then, by construction, we will achieve a quantum performance analogous to the violations of the CGLMP inequalities. We explicitly compute this.

The preparations of Alice are
\begin{equation}
	\rho_{x_0x}=d\Tr_A\left(A_{x_0}^x\otimes \textbf{1}\rho^{AB}\right).
\end{equation}
These can be expanded to 
\begin{equation}
\rho_{x_0x_1}=\frac{1}{N}\sum_{j,k=0}^{d-1}\gamma_j\gamma_k\omega^{(k-j)(x_0-\delta_{x,1}+\alpha_{x})}|k\rangle\langle j|,
\end{equation}
where we have additionally let $x_0\rightarrow x_0-1$ whenever $x=1$. That is no more than a simple relabeling of Alice.

The probability distribution of Bob's outcome is 
\begin{equation}
p(b|x_0,x,y)=\frac{1}{Nd}\sum_{k,j=0}^{d-1} \gamma_k\gamma_j\omega^{(k-j)(x_0-b+\alpha_{x}+\beta_y-\delta_{x,1})}.
\end{equation}
For the probabilities of our interest, as specified in Eq.\eqref{ex2}, put $b=T_r^q$. The resulting distribution does not depend on $x_0$. The final quantum performance can be written 
\begin{multline}\label{effcglmp}
\langle \$\rangle_{x,y}=
\frac{1}{Nd}\sum_{r=0}^{\lfloor \frac{d}{2}\rfloor -1}\sum_{k,l=0}^{d-1}\gamma_k\gamma_l\left(1-\frac{2r}{d-1}\right)\\
\Bigg[\cos\left(\frac{2\pi}{d}(k-l)\left(\frac{1}{4}+r\right)\right)-\cos\left(\frac{2\pi}{d}(k-l)\left(\frac{3}{4}+r\right)\right)\Bigg].
\end{multline}
Note that if we choose the optimal quantum state, by specifying particular $\gamma_k$, as outlined in Ref.\cite{ZG08}, we wil obtain quantum preparation contextuality analogous to the maximal violation of the CGLMP inequalities. 

For our purpose of demonstrating the preparation contextuality of the maximally mixed state, we let Alice and Bob share the maximally entangled state corresponding to $\forall k: \gamma_k=1$ so that the communicated states of Alice averaged over each of the sets in the partition of Alice's input space is the maximally mixed quantum state. To calculate, in that case, the quantum violation of the preparation noncontextuality inequality in Eq.\eqref{ex2}, we use Eq.\eqref{effcglmp}. Inserting our values of $\gamma_k$ and $N=d$, and using that the pair $(k,l)$ only appears as $k-l$ in the series in Eq.\eqref{effcglmp}, along with some trigonometric manipulations, we obtain
\begin{multline}\label{effcmax}
\langle \$\rangle_{x,y}=
\frac{1}{2d^2}\sum_{r=0}^{\lfloor \frac{d}{2}\rfloor -1}\left(1-\frac{2r}{d-1}\right)\\
\Bigg(\csc^2\left(\frac{\pi}{d}\left(r+1/4\right)\right)-\csc^2\left(\frac{\pi}{d}\left(r+3/4\right)\right)\Bigg).
\end{multline}
For every $d$, by construction, this is equivalent to the violations of the CGLMP inequalities presented in Ref.\cite{CGLMP02, ZG08}. This follows directly from Eq.\eqref{res}. Nevertheless, for the sake of examplification, we have numerically checked this for $d=2,\ldots, 200$. 

\end{document}